\documentclass[twocolumn,prd, aps, notitlepage, nofootinbib, showpacs, superscriptaddress,longbibliography]{revtex4-1}
\usepackage[utf8]{inputenc}
\usepackage{graphicx,mathtools,xcolor,latexsym,rotating,soul}
\usepackage[T1]{fontenc}
\usepackage{tikz}
\usepackage{comment}
\usepackage{amsmath, amssymb, amsthm, amsfonts,breqn}
\usepackage{multirow}
\usetikzlibrary{decorations.pathmorphing}
\usepackage[colorlinks=true,citecolor=green,urlcolor=blue,linkcolor=blue]{hyperref}
\usepackage{pifont}
\usepackage{gensymb}
\usepackage{enumitem}
\usepackage{mathrsfs}
\usepackage[scr=boondox,  
            cal=esstix]   
           {mathalpha}

\newtheoremstyle{named}{}{}{\itshape}{}{\bfseries}{.}{.5em}{\thmnote{#3 }#1}
\theoremstyle{named}

\usepackage{blindtext}
\newcommand*{\rom}[1]{\expandafter\@slowromancap\romannumeral #1@}

\makeatletter
\let\cat@comma@active\@empty
\makeatother
\allowdisplaybreaks
\begin{document}
\title{A relativistic treatment of accretion disk torques on extreme mass ratio inspirals around spinning black holes}
\author{Abhishek Hegade K. R.}
\email{ah4278@princeton.edu}
\affiliation{Princeton Gravity Initiative, Princeton University, Princeton, NJ 08544, USA}

\author{Charles F. Gammie}
\email{gammie@illinois.edu}
\affiliation{Illinois Center for Advanced Studies of the Universe, Department of Physics, University of Illinois Urbana-Champaign, Urbana, IL 61801, USA}

\author{Nicol\'as Yunes}
\email{nyunes@illinois.edu}
\affiliation{Illinois Center for Advanced Studies of the Universe, Department of Physics, University of Illinois Urbana-Champaign, Urbana, IL 61801, USA}
\begin{abstract}
We model the motion of a small compact object on a nearly circular orbit around a spinning supermassive black hole, which is also interacting with a thin equatorial accretion-disk surrounding the latter, through tools from self-force and Hamiltonian perturbation theory.
We provide an analytical and relativistically-accurate formalism to calculate the rate of energy and angular momentum exchanged at Lindblad resonances. 
We show that strong relativistic effects can potentially cause a reversal in the direction of the torque on the small compact object if the surface density gradient is not too large.
We analytically explore the dependence of the torque reversal location on the spin of the supermassive black hole and demonstrate that the ratio of the reversal location to the innermost stable circular orbit is approximately insensitive to the spin of the supermassive black hole.
Our results show that relativistic torques can be 1--2 order of magnitude larger than the Newtonian torque routinely used in the literature to model disk/small-compact-object interactions close to the supermassive black hole.
Our results highlight the importance of including relativistic effects when modeling environmental effects in extreme mass-ratio inspirals.
\end{abstract}
\maketitle

\allowdisplaybreaks[4]
\section{Introduction}
Extreme mass-ratio inspirals (EMRIs), where a small compact object (SCO) [of mass in the range $10M_{\odot}$--$100 M_{\odot}$] spirals into a supermassive black hole (SMBH) [of mass $10^5M_{\odot}$--$10^7 M_{\odot}$], emit gravitational waves in the sensitivity band of the Laser Interferometer Space Antenna (LISA). 
EMRIs can accumulate $\gtrsim \mathcal{O}(10^5)$ wave-cycles in the LISA band, and thus, they provide detailed information about the spacetime in the close vicinity of the SMBH. The relativistic nature of EMRI orbits and the proximity of the SCO to the SMBH should allow one to probe the mass and spin of the SMBH, conduct tests of general relativity, and potentially probe the environment around the SMBH~\cite{LISA:2022yao,colpi2024lisadefinitionstudyreport,Babak:2017tow,Kocsis_2011,Yunes:2011ws,Speri_2023,duque2025constrainingaccretionphysicsgravitational}.

Accurate inference of the EMRI system parameters from a future gravitational-wave signal requires waveform models that are highly accurate because of the large number of cycles that accumulate in the LISA band~\cite{colpi2024lisadefinitionstudyreport}.
These high-accuracy requirements are being tackled successfully in \textit{vacuum} (i.e.~in the absence of any external matter source around the SMBH) thanks to advances in self-force and black hole perturbation theory~\cite{LISAConsortiumWaveformWorkingGroup:2023arg,BHPToolkit}.
With the modeling of EMRIs in \textit{vacuum} reaching a mature stage, it is now time to assess the impact of environmental effects on the EMRI motion.

Environmental effects have long been expected to be a small perturbation on EMRIs. This is because EMRIs are expected to exist at Gpc distances from Earth, and thus, their gravitational waves are naturally weak. In order to make them detectable, the SCO must be close enough to the SMBH, so that the SCO's velocity is a good fraction of the speed of light and the gravitational-wave amplitude is not suppressed. The matter-energy density in the vicinity of SMBHs is expected to be small relative to the matter-energy of the SMBH, and thus, any (non-vacuum) environmental effects may be thought to be negligible. 
However, the large number of gravitational-wave cycles EMRIs will deposit in the LISA band will allow us to measure small perturbations from the environment, if the dephasings caused are on the order of a few radians or higher~\cite{Yunes:2011ws,Kocsis_2011}. 

A large body of work has now qualitatively assessed how different environmental effects could be potentially studied with EMRI signals.
A non-exhaustive list of astrophysical environmental effects include the interaction between the SCO and the accretion disk of an SMBH~\cite{2000ApJ...536..663N,2007MNRAS.374..515L,Kocsis_2011,Yunes:2011ws,Speri_2023,duque2025constrainingaccretionphysicsgravitational,Derdzinski:2018qzv,Derdzinski:2020wlw}, mass accretion on to the SMBH and the SCO~\cite{Kocsis_2011,Yunes:2011ws}, viscous drag on the SCO due to gas surrounding the SCO~\cite{Barausse:2007dy}, disk-SCO collisions~\cite{Franchini_2023}, and perturbations due to the presence of a third body~\cite{Yunes:2010sm,Naoz_2013,Gupta_2021,Silva:2022blb,Silva:2025lkl}.
In addition, there could be scenarios where the SCO interact with bosonic clouds and dark matter particles surrounding a SMBH~\cite{Baumann_2022,Maselli_2020,Zhang_2020,Arvanitaki_2011,Vicente_2022,dyson2025environmentaleffectsextrememass}, or where EMRIs are part of a three-body system~\cite{Yunes:2010sm,santos2025gravitationalwavesbemrisdoppler,Yin_2025,cocco2025stronggravityprecessionresonancesbinary,Camilloni_2024}.

If the SMBH is actively accreting, then the dominant source of environmental effect close to the SMBH is the energy and angular momentum exchanged due to the interaction of the SCO with the accretion disk of the SMBH.
Such disk-SCO interactions are reminiscent of the classical disk-satellite interactions studied extensively in the proto-planetary literature~\cite{GT-disc-satellite-interaction,GT-Uranus-Ring,Baruteau_2014,Paardekooper:2023,Murray-Dermott-Book,Ward-1986,2002ApJ...565.1257T,2002ApJ...565.1257T}.
While the interactions are superficially similar, it is essential to recognize that the SCO is moving in the strong-relativistic potential of the SMBH at an appreciable fraction of the speed of light.
Moreover, the properties of an accretion disk close to a SMBH are significantly different from those of a proto-planetary disk. Although a precise understanding of properties of accretion disks close to a SMBH is lacking, some important properties, such as magnetization, turbulence, and radiation from the plasma, are expected to play an important role (see~\cite{Blaes:2014,Davis:2020,White:2023} for recent discussion and progress). 
Hence, any realistic model of a disk-SCO interaction close to a SMBH should include relativistic effects and model the microphysics of the accretion disk.

Several studies have assessed the potential for EMRIs to probe disk-SCO interactions~\cite{Yunes:2011ws,Kocsis_2011,Sberna:2022qbn,Speri_2023,Copparoni_2025,duque2025constrainingaccretionphysicsgravitational,Derdzinski:2018qzv,Derdzinski:2020wlw}; however, except for~\cite{Hirata_2011,Hirata_2011_II}, all of these studies have used the disk-satellite formula derived in the proto-planetary literature to model the disk-SCO interaction.
In addition, the relativistic model introduced in~\cite{Hirata_2011,Hirata_2011_II} (and studied more recently in~\cite{future-work-in-prep-Duque}) is a purely numerical approach that relies on the integration of the Teukolsky equation.
In an effort to build simple relativistic models of disk-SCO interaction close to a SMBH, we introduced an analytical model in~\cite{HegadeKR:2025dur} (Paper I, hereafter), using tools from self-force and Hamiltonian perturbation theory.
The methods introduced in Paper I model the interaction between the disk and the SCO as a three-body problem, where a disk particle close to the orbit of the SCO exchanges energy and angular momentum with the SCO at Lindblad resonances\footnote{
Lindblad resonances occur when the epicylic frequency of the SCO orbit is an integer multiple of the forcing frequency of the density waves raised in the accretion disk. Such resonances are the dominant source of angular and momentum exchange in disk-SCO interactions, see~\cite{Binney-Tremaine-Book,GT-disc-satellite-interaction} for a detailed discussion.}.
Such simplified methods were used to study Newtonian proto-planetary disks in the past~\cite{GT-disc-satellite-interaction,GT-Uranus-Ring,Ogilvie_2007,Ward-1986,Ward-1997}, and our approach extended these methods to a Schwarzschild (non-rotating black hole) background setting.

Although our analytical model ignores the microscopic properties of the disk, it can capture the properties of the gravitational interaction between the disk and the SCO. Indeed, we identified two important relativistic effects in disk-SCO interactions around non-spinning SMBHs that are not present in Newtonian gravity: 
\begin{itemize}
    \item[i)] A change in the distribution of the inner and outer Lindblad resonances close to the orbit of the SCO due to strong relativistic effects.
    \item [ii)] A potential reversal in the direction of the torque on the SCO due to the disk if the density gradients are not too large.
\end{itemize}
These effects have been confirmed by an independent work~\cite{future-work-in-prep-Duque} that follows the approach of~\cite{Hirata_2011,Hirata_2011_II}. 

In this paper, we employ the methods introduced in Paper I to investigate the impact of the SMBH spin on the disk-SCO interaction.
We provide analytic formula for the torque and power exchanged due to disk-SCO interactions and explore the impact of the spin of the SBMH on disk-SCO interactions. 
We show that, for retrograde disk and SCO orbits, the torque reversal can occur at distances as large as 7.5 Schwarzschild radii from the SMBH.
This effect arises because the ISCO for retrograde orbits progressively moves outward, as the spin of the SMBH increases.
We also show that the ratio of the torque reversal location to the location of the innermost stable circular orbit (ISCO) is approximately insensitive to the spin of the SMBH.
However, the torque reversal location can be affected by strong density gradients in the disk.

The rest of the paper explains our results in detail and is organized as follows:
In Sec.~\ref{sec:qualitative-Newt-Intro}, we extrapolate results from Newtonian theory by substituting the relativistic epicyclic frequencies to qualitatively show how torque reversal arises.
In Sec.~\ref{sec:energy-and-angular-momentum-exchange}, we extend the analysis of Paper I to include the spin of the SMBH and provide analytic formula for the rate of energy exchanged at Lindblad resonances.
In Sec.~\ref{sec:results}, we use a simple analytical model for the disk surface density to explore how the torque reversal location depends on the spin of the SMBH and compare the disk torque to torques from Newtonian theory and gravitational-wave emission.
Our conclusions are presented in Sec.~\ref{sec:conclusions}.
A review of orbital mechanics in a Kerr background spacetime using action-angle variables is presented in Appendix~\ref{appendix:orbital-mechanics-Kerr}. Henceforth, we use geometric units, with $G=1=c$.

\section{Qualitative discussion of the impact of spin on torque reversal}\label{sec:qualitative-Newt-Intro}
As in Sec.~II of Paper I, we here qualitatively describe the impact of spin on the torque imparted by the disk on the SCO by extrapolating the Newtonian formula derived in the proto-planetary literature~\cite{Ward-1986}.
Our relativistically-accurate results are presented in Sec.~\ref{sec:energy-and-angular-momentum-exchange}, but the Newtonian presentation here provides some of the key insights.

Consider a SCO of mass $m$ on an equatorial, circular orbit around a SMBH of mass $M$ at radius $r = p' M$\footnote{A prime denotes the orbital elements of the SCO, unless otherwise stated.} 
in Boyer-Lindquist coordinates that interacts with a thin, equatorial disk of surface density $\Sigma(r)$.
We assume that the angular momentum of the disk and the SCO point in the same direction, but we let the angular momentum of the SMBH be arbitrary.
The dominant source of torque for the disk-SCO configuration described above arises from the exchange of energy and angular momentum at Lindblad resonances~\cite{GT-disc-satellite-interaction}.
The location of Lindblad resonances $r_{\mathrm{Lr}} = p_{\mathrm{Lr}} M$ in Newtonian gravity are given by the roots of~\cite{GT-disc-satellite-interaction,Ward-1986}
\begin{align}\label{eq:GT-formula-location-of-LR-Resonances}
    \Omega(p_{\mathrm{Lr}}) - \Omega(p') &= - \frac{k}{j} \kappa (p_{\mathrm{Lr}})
\end{align}
where $\Omega$ is the angular velocity of the SCO, $\kappa$ is the epicyclic frequency of the orbit, $k=\pm 1$ denotes whether the resonance is inside $(k=-1)$ or outside $(k=1)$ the orbit of the SCO, and $j$ is the mode-number of the resonances, which is approximately equal to the inverse of the aspect ratio of the disk.

The orbital frequency and the epicyclic frequencies in a Kerr spacetime as measured by an observer at infinity are given by~\cite{Novikov-Thorne,Abramowicz_2013,Gammie_2004}
\begin{subequations}\label{eq:Omega-and-Kappa-Kerr}
\begin{align}
    \Omega(p) &= \frac{1}{M (p^{3/2} + \chi)} \,,\\
    \kappa(p) &= \Omega(p) 
    \sqrt{1 -\frac{6}{p} -\frac{3 \chi ^2}{p^2} + \frac{8 \chi }{p^{3/2}}}
    \,.
\end{align}
\end{subequations}
where $\chi = a/M$ is the dimensionless spin of the SMBH. 
If $\chi>0$, the SCO and the disk are in a prograde orbit, and when $\chi<0$, the orbits are retrograde. 

We can use the values of the orbital frequency and the epicyclic frequency from Eq.~\eqref{eq:GT-formula-location-of-LR-Resonances} to obtain the approximate locations of the resonances close to the orbit of the SCO (i.e.~those with $j \gg 1$), which dominate the interactions. 
Solving Eq.~\eqref{eq:GT-formula-location-of-LR-Resonances} in the large $j$ limit, one finds
\begin{align}\label{eq:Lindblad-location-intro}
    &p_{\mathrm{Lr}} \!\!=\!\! p' 
    \Bigg[1
    +
    \frac{2 k \left(\left(p'\right)^{3/2}+\chi \right)}{3 (p')^{3/2} j}
    \sqrt{1 -\frac{6}{p} -\frac{3 \chi ^2}{p^2} + \frac{8 \chi }{p^{3/2}}}
    \nonumber\\
    &-\frac{\left(\left(p'\right)^{3/2}+\chi \right)^2 \left(32 \chi  \sqrt{p'}+\left(p'\right)^2-18 p'-15 \chi ^2\right)}{9 \left(p'\right)^5 j^2}
    \Bigg]
    \nonumber\\
    &+
    \mathcal{O}(j^{-3})\,.
\end{align}
Observe that the magnitude of the coefficient of the $\mathcal{O}(j^{-1})$ term is the same for the inner and the outer resonances. 
This tells us that the inner ($k=-1$) and outer ($k=1$) Lindblad resonances are symmetrically distributed around the location of the SCO to $\mathcal{O}(j^{-1})$.

\begin{figure}
    \centering
    \includegraphics[width=1\linewidth]{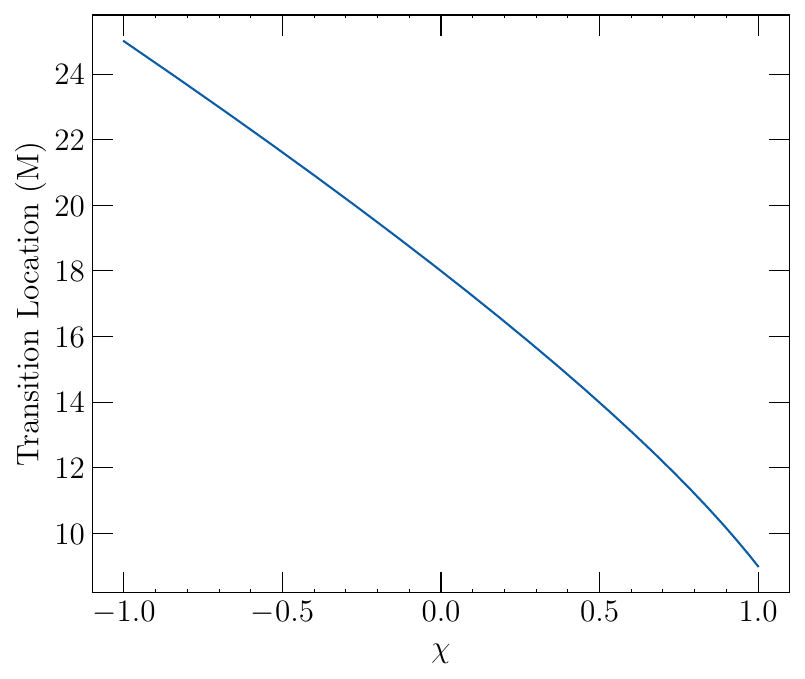}
    \caption{Location where the inner Lindblad resonances begin to get closer to the orbit of the SCO as a function of the spin of the SMBH. 
    Observe that the transition occurs in the inner regions of the SMBH where EMRI are expected to be detected in the LISA band.}
    \label{fig:transition-Newt}
\end{figure}

The asymmetry in the location of resonances arises from $\mathcal{O}(j^{-2})$ term. 
The sign of this term depends on the sign of 
\begin{align}
    \mathcal{I}(p',\chi)
    \equiv 
    -\frac{32 \chi  \sqrt{p'}+\left(p'\right)^2-18 p'-15 \chi ^2}{(p')^2}
    \,.
\end{align}
If $\mathcal{I}<0$ then, the outer Lindblad resonances are closer to the orbit of the SCO than the inner Lindblad resonances, and this is precisely what happens in the Newtonian limit, where $\mathcal{I}_{\mathrm{Newt}} = -1$.
Closer to the SMBH, one could potentially have $\mathcal{I}>0$, implying that the inner Lindblad resonances are closer to the SCO. 
In Paper I, we showed that for a Schwarzschild SMBH ($\chi=0$), the inner Lindblad resonances are closer than the outer Lindblad resonance if $p'<18$.
The dependence of the transition location [$\mathcal{I}(p'_{\mathrm{trans}},\chi) = 0$] on the spin of the SMBH is shown in Fig.~\ref{fig:transition-Newt}.
Observe that the location depends sensitively on the spin of the SMBH, and if the orbit is retrograde and the SMBH is maximally spinning, then the location can be as large as $p' = 25$.

The total differential torque on the SCO can be obtained by using the expressions from~\cite{Ward-1986}. To qualitatively understand the change in the sign of the torque due to relativistic effects, we proceed as in Paper I and assume that the inner and outer torque cutoff parameters are the same; then, the torque is given by~\cite{Ward-1986}
\begin{align}\label{eq:LR-torque-Newt-Intro}
    &\left< \dot{L}'_{z} \right>_{\mathrm{approx}}
    \propto
    \nonumber\\
    &
    \frac{4 q^2 M^2 \Sigma(p') \mathcal{G}\left(p'\right) }{F_0 \Omega(p')^2 \Omega_1^4 (p')^2}
    \left[2 K_0\left(\frac{F_0}{\Omega_1}\right)
    + F_0 K_1\left(\frac{F_0}{\Omega_1}\right) \right]  
    \nonumber\\
    &-
    \frac{d \Sigma}{dp'}
    \frac{4 q^2 M^2}{F_0 \Omega(p')^6 \Omega_1^4 p'}
    \left[2 K_0\left(\frac{F_0}{\Omega_1}\right)
    + F_0 K_1\left(\frac{F_0}{\Omega_1}\right) \right]^2
    \,,
\end{align}
where $q=m/M \ll 1$ is the mass ratio, we have ignored a positive proportionality constant related to the torque cutoff parameters, $\dot{L}_{z}'$ is the rate of change of the angular momentum of the SCO in the $z$-direction, the angular brackets denote an average over the orbital period of the SCO, $F_0 \equiv \kappa(p')/\Omega(p')$, and $K_{0,1}$ are modified Bessel functions. 
The subscript $\mathrm{approx}$ highlights the fact that the above equation is an approximation that extrapolates the values of the epicyclic frequencies from Newtonian gravity to general relativity. The fully relativistic expressions for the differential Lindblad torque will be derived in Sec.~\ref{sec:Lindblad-resonances}.
The function $\mathcal{G}(p')$ is discussed below and we have defined 
\begin{align}\label{eq:Omega1Omega2def}
    \Omega_1 \equiv - \frac{d \log (\Omega(p'))}{d \log (p')}\,, \Omega_2 \equiv - \frac{d^2 \log (\Omega(p'))}{d^2 \log (p')}\,.
\end{align}

By inspecting Eq.~\eqref{eq:LR-torque-Newt-Intro}, we see that the arguments of the Bessel functions depend on $F_0 \propto \kappa$. 
The epicyclic frequency is zero at the ISCO, implying that the modified Bessel functions diverge there.
This divergence qualitatively explains why the magnitude of the torque should increase as the SCO approaches the ISCO.
The torque reversal phenomena can also be qualitatively understood by examining Eq.~\eqref{eq:LR-torque-Newt-Intro}. 
The term proportional to $d\Sigma/dp'$ is always negative and one can show that, unless the density gradients are large, this term is subdominant to the term proportional to $\Sigma(p')$ in Eq.~\eqref{eq:LR-torque-Newt-Intro}.
The term proportional to $\Sigma(p')$ can be positive or negative depending on the sign of
\begin{align}\label{eq:G-def}
    &\mathcal{G}(p')
    \equiv
    K_0\left(\frac{F_0}{\Omega_1}\right)
    \left[
    \frac{\Omega_1 \left(-4 (-1+\kappa_1) \Omega_1+\Omega_1^2-2 \Omega_2\right)}{F_0}
    \right. \nonumber\\
    &\left. 
    +F_0 \left(-((1+2 \kappa_1) \Omega_1)+\Omega_1^2+\Omega_2\right)
    \right]
    \nonumber\\
    &+
    K_1\left(\frac{F_0}{\Omega_1}\right)
    \left[ -2 (1+2 \kappa_1) \Omega_1+(3-2 \kappa_1) \Omega_1^2+2 \Omega_2\right]
    \,.
\end{align}
where $\kappa_1 \equiv - d \log (\kappa(p'))/d \log (p')$.
\begin{figure}
    \centering
    \includegraphics[width=1\linewidth]{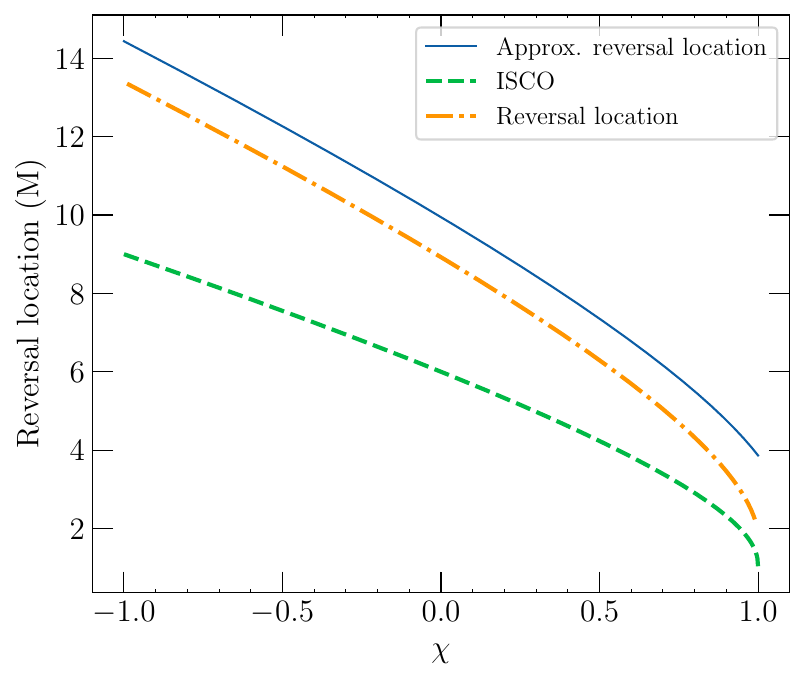}
    \caption{The torque reversal location as a function of the spin of the SMBH.  The solid blue curve shows the approximate location where torque changes sign as a function of the spin of the SMBH obtained by solving $\mathcal{G}(p') = 0$ [Eq.~\eqref{eq:G-def}].
    The orange dash-dotted curve is the relativistically accurate location obtained by solving $\mathcal{G}_1(p')$ [Eq.~\eqref{eq:G1-location}].
    The green dashed curve shows the location of the ISCO.}
    \label{fig:location-of-reversal}
\end{figure}
In Paper I, we showed that for SCOs around non-rotating black holes (i.e.~in a Schwarzschild background spacetime), $\mathcal{G}(p')$ is negative only if $p'>9.94$, signaling a change in the sign of the torque at this separation. 
Figure~\ref{fig:location-of-reversal} shows how the location of the torque reversal $p'_{\mathrm{rev\,from\,Newt}}$ depends on the spin of the SMBH. The solid blue curve indicates the approximate location derived in this section, while the relativistically accurate location is depicted as a dash-dotted orange curve, as obtained in Sec.~\ref{sec:reversal}. The green dashed curve represents the location of the ISCO.
The extrapolation used here slightly overestimates the torque reversal location, but it provides a reasonable order-of-magnitude estimate. We also see that the ratio of the reversal location to the ISCO location is approximately constant with SMBH spin, a fact we will discuss in more detail in the next sections.

\section{Energy and angular momentum exchanged due to disk-SCO interactions}\label{sec:energy-and-angular-momentum-exchange}

The formalism for computing the power and torque imparted on the SCO due to disk-SCO interaction was described in detail in Sec. III of Paper I. 
In Sec.~\ref{sec:review-formalism} below, we briefly review the key elements of our formalism for completeness and refer the reader to Paper I for further details. 
Next, we concentrate on Lindblad resonances and provide an analytic formula for the rate of energy and angular momentum exchange in Sec .~\ref {sec:Lindblad-resonances}.

\subsection{Review of formalism to compute energy and angular momentum exchange}\label{sec:review-formalism}
In Sec.~III of Paper I, we studied the disk-SCO interaction by first modeling the local gravitational interaction between the SCO and a small particle element of mass $d\mu$ in the disk. We then integrated over the entire disk to obtain the total contribution to the disc-SCO interaction. 
The formalism can be split into the following three steps:
\begin{itemize}
    \item \textit{Step 1}: Modeling the interaction between the disk element and the SCO using the singular potentials of the worldline.
    \item \textit{Step 2}: Obtaining the Hamiltonian for the disk-SCO system in action-angle variables.
    \item \textit{Step 3}: Using Hamiltonian perturbation theory to obtain the power and torque on the SCO and the disk.
\end{itemize}
We now briefly review each of these steps and refer the reader to Sec.~III of Paper I for further technical details.

\subsubsection{Step 1: SCO-Disk Interaction}
Let us assume that the disk mass element moves on a worldline $z^{\mu}$ and the SCO moves on worldline $z^{\mu'}$.
The motion of the disk element can be obtained by varying the point-particle action
\begin{align}\label{eq:action-variable}
    S = -\int dt \sqrt{-g_{\mu \nu} \dot{z}^{\mu} \dot{z}^{\nu}}\,,
\end{align}
where the metric is given by
\begin{align}
    g_{\mu \nu} = g_{\mu \nu}^{\mathrm{Kerr}} + \delta g_{\mu\nu}
\end{align}
and $g^{\mathrm{Kerr}}$ is the Kerr metric of the SMBH, while $\delta g$ is the contribution due to the SCO.
As we described in Eq.~(7) of Paper I, $\delta g_{\mu \nu}$ can be separated into a sum of three different terms:
\begin{align}\label{eq:metric-pert-split-v1}
    \delta g_{\mu \nu}(x) = h_{\mu\nu}^{\mathrm{S},d\mu}(x;z) + h_{\mu\nu}^{\mathrm{R},d\mu}(x;z) + h_{\mu \nu}^{\mathrm{int}}(x;z') \,,
\end{align}
where $h_{\mu\nu}^{\mathrm{S},d\mu}(x;z)$ is the singular field of the mass element of the disk, $h_{\mu\nu}^{\mathrm{R},d\mu}(x;z)$ is the regular field of the mass element of the disk, and $h_{\mu \nu}^{\mathrm{int}}(x;z')$ is an interaction term that is generated by the SCO.
The singular and the regular field of the disk's mass element are its self-field and its radiative field.
The self-field does not contribute to the motion of the disk element, while the radiative field only induces a radiative force on it. These contributions can be ignored when studying disk-SCO interactions. 

Therefore, when evaluated in a close neighborhood of the worldline of the disk element, the perturbation to the metric can be approximated as 
\begin{align}\label{eq:perturb-metric}
    \delta g_{\mu \nu}(z) &\approx h_{\mu \nu}^{\mathrm{int}}(z;z') = h_{\mu\nu}^{\mathrm{S,SCO}}(z;z') + h_{\mu\nu}^{\mathrm{R,SCO}}(z;z')
    \nonumber\\
    &\approx
    h_{\mu\nu}^{\mathrm{S,SCO}}(z;z')\,,
\end{align}
where $h_{\mu\nu}^{\mathrm{S,SCO}}(z;z')$ and $h_{\mu\nu}^{\mathrm{R,SCO}}(z;z')$ are the singular and the regular field contributions from the SCO.
In the second line of the above equation, we have assumed that the Newtonian-like singular potential of the SCO $h_{\mu\nu}^{\mathrm{S,SCO}}(z;z')$ is the dominant contribution to the interaction, and we ignored the regular field of the SCO by assuming that the resonant interaction is quasi-adiabatic.
Moreover, the regular field of the SCO will be a very high post-Newtonian contribution to the dynamics, which can be incorporated in our analysis by using numerical self-force calculations~\cite{BHPToolkit}.
With this assumption, we can use analytic expressions for the singular field~\cite{Pound_2014,Heffernan_2012} to evaluate the above equation (see Paper I for further details).

\subsubsection{Step 2: Disk-SCO Hamiltonian}
With the perturbed metric in hand [Eq.~\eqref{eq:perturb-metric}], we can use the Lagrangian of Eq.~\eqref{eq:action-variable} to obtain the Hamiltonian $H$ for the disk mass element. 
Hamilton's equations for the disk-SCO system can be written in the following form [see e.g.~Eq.~(21) of Paper I]:
\begin{subequations}\label{eq:hamiltonian-equations-3-body}
\begin{align}
    \dot{\mathbf{P}}_{i} &= \epsilon_{\mathrm{int}}m \frac{\partial \mathcal{R}}{\partial \mathbf{Q}^{i}} \,,\\
    \dot{\mathbf{Q}}^{i} &= \frac{\partial H}{\partial \mathbf{P}_{i}} - \epsilon_{\mathrm{int}}m\frac{\partial \mathcal{R}}{\partial \mathbf{P}^{i}} \,,\\
    \dot{\mathbf{P}}'_{i} &= \epsilon_{\mathrm{int}}\int d \mu \frac{\partial \mathcal{R}'}{\partial \mathbf{Q}^{'i}} \,,\\
    \dot{\mathbf{Q}}^{'i} &= \frac{\partial H'}{\partial \mathbf{P}_{i'}} - \epsilon_{\mathrm{int}}\int d \mu \frac{\partial \mathcal{R'}}{\partial \mathbf{P}'_{i}} \,,
\end{align}
\end{subequations}
where the over dots denote time derivatives, $(\mathbf{P}_{i},\mathbf{Q}^{i})$ are any set of action-angle variables for the disk mass-element, $(\mathbf{P}'_{i},\mathbf{Q}^{'i})$ are a set of general action-angle variables for the SCO, 
$(\mathcal{R},\mathcal{R}')$ are the disturbing functions,
and $\epsilon_{\mathrm{int}}$ is an order counting parameter that is used to perturbed the geodesic orbit.
The disturbing function $\mathcal{R}$ quantifies the perturbation caused on the disk element due to the SCO, while $\mathcal{R}'$ quantifies the perturbation on the orbit of the SCO due to the 
disk element.
In the Newtonian limit, the disturbing functions are the Newtonian field of the disk and the SCO, and they have exactly the same form since the mass has been scaled out in Eq.~\eqref{eq:hamiltonian-equations-3-body}, namely
\begin{align}
    \mathcal{R}_{\mathrm{Newt}} = \mathcal{R}'_{\mathrm{Newt}} 
    \propto
    \frac{1}{|z - z'|}
    \,,
\end{align}
while relativistic expressions can be found in Eq.~(19) of Paper I. 
Equation~\eqref{eq:hamiltonian-equations-3-body} is valid for any set of action-angle variables, but we here use $\left(\left\{\Lambda,\lambda\right\},\left\{P,-\varpi\right\}\right)$, as defined in Eq.~\eqref{eq:modified-Delaunay-variables}.
We also expand the disturbing function in a Fourier series using the angle variables $(\lambda,-\varpi)$. 
For small eccentricities, the Fourier series can be written down schematically as [see Eqs.~(49) and (50) of Paper I]
\begin{subequations}\label{eq:R-dist-function-expr-lin-e}
\begin{align}
    \mathcal{R} &= 
    \frac{1}{2}
    \sum_{j=-\infty}^{\infty} 
    \Bigg[R_{(-j,0,0),(j,0,0)} \cos(j\lambda'- j\lambda)
    \nonumber\\
    &+
    R _{(-j-1,-1,0),(j,0,0)} \cos(j\lambda' - (j+1) \lambda  + \varpi)
    \nonumber\\
    &+
    R_{(-j+1,1,0),(j,0,0)} \cos(j\lambda' -(j-1) \lambda - \varpi)
    \nonumber\\
    &+
    R _{(-j-1,0,0),(j,-1,0)} \cos(j \lambda' -(j+1) \lambda + \varpi')
    \nonumber\\
    &+
    R_{(-j+1,0,0),(j,1,0)} \cos(j \lambda' -(j-1) \lambda - \varpi')
    \Bigg]
    \nonumber\\
    &+
    \mathcal{O}(e^2,(e')^2,ee'), \\
    \mathcal{R}' &= 
    \frac{1}{2}
    \sum_{j=-\infty}^{\infty}  
    \Bigg[R'_{(-j,0,0),(j,0,0)} \cos(j\lambda'- j\lambda)
    \nonumber\\
    &
    +
    R' _{(-j-1,-1,0),(j,0,0)} \cos(j\lambda' - (j+1) \lambda  + \varpi)
    \nonumber\\
    &+
    R'_{(-j+1,1,0),(j,0,0)} \cos(j\lambda' -(j-1) \lambda - \varpi)
    \nonumber\\
    &+
    R' _{(-j-1,0,0),(j,-1,0)} \cos(j \lambda' -(j+1) \lambda + \varpi')
    \nonumber\\
    &+
    R'_{(-j+1,0,0),(j,1,0)} \cos(j \lambda' -(j-1) \lambda - \varpi')
    \Bigg]
    \nonumber\\
    &+
    \mathcal{O}(e^2,(e')^2,ee'),
\end{align}
\end{subequations}
where $e$ and $e'$ are the eccentricities of the orbit of the disk-element and of the SCO (both around the SMBH), while the coefficients of the series satisfy
\begin{subequations}\label{eq:scaling-relation-dist-func}
\begin{align}
    &R_{(-j,0,0),(j,0,0)}, R'_{(-j,0,0),(j,0,0)} = \mathcal{O}(1)   \,,\\
    &R_{(-j,0,0),(j,0,0)}, R'_{(-j,0,0),(j,0,0)} = \mathcal{O}(1) \,,\\
    &R_{(-j-1,-1,0),(j,0,0)},R' _{(-j-1,-1,0),(j,0,0)} = \mathcal{O}(e) \,,\\
    &R_{(-j+1,1,0),(j,0,0)},R'_{(-j+1,1,0),(j,0,0)} = \mathcal{O}(e) \,,\\
\label{eq:dist-for-corotation-1}
    &R_{(-j-1,0,0),(j,-1,0)}, R'_{(-j-1,0,0),(j,-1,0)} = \mathcal{O}(e')\,,\\
\label{eq:dist-for-corotation-2}
    &R_{(-j+1,0,0),(j,1,0)}, R'_{(-j+1,0,0),(j,1,0)} = \mathcal{O}(e')\,.
\end{align}
\end{subequations}
The explicit expressions for the Fourier coefficients of Eq.~\eqref{eq:R-dist-function-expr-lin-e}, such as $R_{(-j,0,0),(j,0,0)}$, were derived for a Schwarzschild background spacetime in Paper I [see Eq.~(58) and the supplementary material for Paper I]. 
We have extended this calculation to a Kerr background spacetime, and these expressions are provided in the supplementary \texttt{MATHEMATICA} notebook. 

\subsubsection{Step 3: Power and Torque from Hamiltonian Perturbation Theory}

In this step, we solve the Hamilton equations [see Eq.~\eqref{eq:hamiltonian-equations-3-body}] perturbatively in $\epsilon_{\mathrm{int}}$.
At first order in perturbation theory, the actions (energy and angular momentum) evolve periodically, and when averaged over the orbit of the SCO and the disk, these perturbations vanish.
At second-order in perturbation theory, the actions evolve on a secular timescale, leading to a non-zero torque and energy flux.
The technique used to obtain the perturbative results is a standard tool in Hamiltonian mechanics~\cite{K-Cole-perturbation-theory,GT-Uranus-Ring}.
We describe the procedure in detail in Sec. III of Paper I and the final result for the secular evolution of the action variables are given in Eqs.~(36) and (37) of Paper I.

\subsection{Lindblad resonances}\label{sec:Lindblad-resonances}
Corotation and Lindblad resonances occur when the arguments of the disturbing function are in resonance. In particular, Lindblad resonances arise when
\begin{align}
    \label{eq:Lindblad-resonance-condition}
    \dot{\Phi}_{\mathrm{Lr}} &= j \dot{\lambda} ' - (j + k) \dot{\lambda} + k \dot{\varpi} = 0\,,\quad  k = \pm 1.
\end{align}
Using the expression for the frequencies of the system from Appendix~\ref{appendix:small-eccentricity}, one can show that the locations of the Lindblad resonances are exactly as presented in Eq.~\eqref{eq:Lindblad-location-intro}.

The energy and angular momentum exchanged due to Lindblad resonances can be obtained using the secular evolution equation derived in Paper I and the expressions for the disturbing function in a Kerr background spacetime derived in this paper. The final expressions for the $k=\pm 1$ resonance can be written schematically as
\begin{subequations}\label{eq:dot-E-L-Lr-eqns}
\begin{align}
    &\left<\dot{E}'\right>_{\mathrm{Lr},k} = 
    j^2  \left< \dot{E}' \right>_{\mathrm{Lr,k}}^{(0)}
    +
    j \left< \dot{E}' \right>_{\mathrm{Lr,k}}^{(1)}
    +
    \mathcal{O}(1)
    \,,\\
    \label{eq:Lindblad-torque-k-1}
    &\left< \dot{L}'_{z}\right>_{\mathrm{Lr}} = 
    j^2  \left< \dot{L}'_z \right>_{\mathrm{Lr,k}}^{(0)}
    +
    j \left< \dot{L}'_z \right>_{\mathrm{Lr,k}}^{(1)}
    +
    \mathcal{O}(1)
    \,,
\end{align}
\end{subequations}
where $j$ is the mode number. The expressions for $\left< \dot{E}' \right>_{\mathrm{Lr,k}}^{(0,1)}$ and $\left< \dot{L}'_z \right>_{\mathrm{Lr,k}}^{(0,1)}$ in the Newtonian limit and in a Schwarzschild background spacetime were provided in Eqs.~(80)--(82) of Paper I. 
The expressions in a Kerr background spacetime are given by
\begin{widetext}
\begin{subequations}\label{eq:Lindblad-E-expr-actual}
\begin{align}
    &\left< \dot{E}' \right>_{\mathrm{Lr,k}}^{(0)}
    =
    -\frac{16 M q^2 k x^{26}\Sigma[p'] \left(2 \chi +x^6-3 x^2\right)^4}{3 \left(\chi +x^6\right)^4 \left(-3 \chi ^2+x^8-6 x^4+8 \chi  x^2\right)^{3/2}}
    \bigg[ 
    K_0(X)+
    b_{0} K_1(X) 
    \bigg]^2
    \left(\chi ^2+x^8-2 x^4\right)^{-2}
    \,,\\
\label{eq:Edot-Relativistic-expr-LR}
    &\left< \dot{E}' \right>_{\mathrm{Lr,k}}^{(1)}
    =
    -\frac{32 M q^2 x^{20} \left(2 \chi +x^6-3 x^2\right)^4 d\Sigma[p']/dp'}{9 \left(\chi +x^6\right)^3 \left(-3 \chi ^2+x^8-6 x^4+8 \chi  x^2\right) } 
    \bigg[ 
    K_0(X)+
    b_0 K_1(X) 
    \bigg]^2
    \left(\chi ^2+x^8-2 x^4\right)^{-2}
    \nonumber\\
    &-
    \frac{8 M b_{1} q^2}{27}  \Sigma(p')
    \bigg[ 
    K_0(X)+
    b_0 K_1(X) 
    \bigg]
    \bigg[ 
    K_0\left(X\right)
    +
    b_2
    K_1\left(X\right)
    \bigg]
    \,.
\end{align}
\end{subequations}
\end{widetext}
where
\begin{subequations}
\begin{align}
    x &\equiv (p')^{1/4} \,,\\
    X &\equiv  \frac{2 x \sqrt{2 \chi +x^6-3 x^2} \sqrt{-3 \chi ^2+x^8-6 x^4+8 \chi  x^2}}{3 \left(\chi ^2+x^8-2 x^4\right)}\,, \\
    b_0 &\equiv \frac{\sqrt{-3 \chi ^2+x^8-6 x^4+8 \chi  x^2}}{2 x \sqrt{2 \chi +x^6-3 x^2}} \,,
\end{align}
\end{subequations}
and the coefficients $b_{1}$ and $b_2$ are listed in Appendix~\ref{appendix:b-12}.
Note that $X = \kappa/2|A|$ where $|A|$, is the local shear rate is the Fermi frame~\cite{Gammie_2004}.
The expressions for $\left< \dot{L}'_z \right>_{\mathrm{Lr,k}}^{(0,1)}$ can be derived from Eq.~\eqref{eq:Lindblad-E-expr-actual} using that
\begin{align}
\label{eq:Ldot-Relativistic-expr-LR}
    &\left< \dot{L}'_z \right>_{\mathrm{Lr,k}}^{(0,1)}
    =
    \frac{1}{\Omega(p')} 
    \left< \dot{E}' \right>_{\mathrm{Lr,k}}^{(0,1)}
    \,
\end{align}
for a circular orbit.
We note that the value of $\left< \dot{E}' \right>_{\mathrm{Lr,k}}^{(0)}$ can also be obtained by a calculation in the local Fermi frame, see Sec. IV C of Paper I and this is a consistency check on our calculation.

The total secular rate of change of orbital elements is obtained by summing over all the inner ($k=-1$) and outer ($k=1$) Lindblad resonances 
\begin{align}\label{eq:lindblad-resonance-total}
    \left<\dot{E}'\right>_{\mathrm{Lr,tot}} 
    &= 
    \sum_{j\gg 1}
    \left<\dot{E}'\right>_{\mathrm{Lr,+}} + \left<\dot{E}'\right>_{\mathrm{Lr,-}}
    \nonumber \\
    &= 
    \bigg(\frac{j_{\mathrm{max,Lr,+}}^3}{3} - \frac{j_{\mathrm{max,Lr,-}}^3}{3}\bigg)\left< \dot{E}' \right>_{\mathrm{Lr,+}}^{(0)}
    \nonumber\\
    &
    +
    \frac{j_{\mathrm{max,Lr,+}}^2}{2}\left< \dot{E}' \right>_{\mathrm{Lr,+}}^{(1)} 
    + \frac{j_{\mathrm{max,Lr,-}}^2}{2}\left< \dot{E}' \right>_{\mathrm{Lr,-}}^{(1)}
    \,,
\end{align}
where we used the asymptotic approximation $\sum_{j} j^n \approx j_{\mathrm{max,Lr,\pm}}^{n+1}/(n+1)$, where $j_{\mathrm{max,Lr,\pm}}$ are cutoff parameters that regulate the location of the inner and outer Lindblad resonances. Analogous expressions hold for the angular momentum flux $\left<\dot{L}'_{z}\right>$.
Equation~\eqref{eq:Edot-Relativistic-expr-LR} is one of the main results of the paper. 
To the best of our knowledge, this is the first time the expressions for the power exchanged due to disk-SCO interactions have been calculated for motion in a Kerr background spacetime.
These equations generalize results obtained in a Schwarzschild background spacetime in Paper I, and they are valid for EMRIs around a Kerr background of any spin. 

\section{Torque Reversal, comparison to Newtonian theory and gravitational wave emission}\label{sec:results}
In this section, we study the consequences of the relativistic torque formula for EMRIs in an accretion disk around a spinning SMBH.
To analyze the relativistic torque formula, we need a model for the disk surface density $\Sigma$ and the torque cutoff parameter $j$. 
As in Paper I, we assume that the disk surface density and the aspect ratio $h(r)$ can be parameterized by a simple power law,
\begin{subequations}\label{eq:disk-parameterization}
\begin{align}
    &\Sigma = \Sigma_{0} \left(\frac{r}{10 M} \right)^{-\Sigma_{p}} \mathrm{g \; cm^{-2}}\,,\qquad
    h = h_0 \left(\frac{r}{10 M} \right)^{-\Sigma_{h}}
    \,,
\end{align}
\end{subequations}
where $\Sigma_{0}, h_0, \Sigma_{p}$ and $\Sigma_{h}$ are constants. 
We also assume that the inner and outer cutoff parameters are equal to the ratio of the orbital velocity of the SCO to the local speed of sound
\begin{align}\label{eq:j-max-ansatz}
    j_{\mathrm{max}}
    &=
    j_{\mathrm{max,Lr,\pm}}
    =
    \left.\frac{r \Omega}{c_{s}}\right|_{r=p'M}
    \approx
    \left.\frac{1}{h}\right|_{r=p'M}
    =
    \frac{1}{h_0} \left(\frac{p'}{10} \right)^{\Sigma_{h}}
    \,.
\end{align}
From hereon, we consider $\Sigma_{0} \in \left[10^3, 10^7\right]$ and $h_{0} \in \left[0.015,0.15\right]$, and we set $\Sigma_{h} = 1$ for simplicity. These ranges are approximately consistent with $\alpha$ and $\beta$ disk-like profiles of accretion disks (see discussion in Paper I).

The rest of this section is organized as follows:
In Sec.~\ref{sec:reversal} we discuss the impact of spin on the torque reversal phenomena.
We then discuss how the relativistic torque formula compares to the Newtonian torque formula in the inner regions of the black hole in Sec.~\ref{sec:comparison-to-Newt}.
Finally, we compare the torque due to disk-SCO interaction with the gravitational wave flux in Sec.~\ref{sec:comparison-to-GW}.
\subsection{Torque reversal}\label{sec:reversal}
\begin{figure}
    \centering
    \includegraphics[width=1\linewidth]{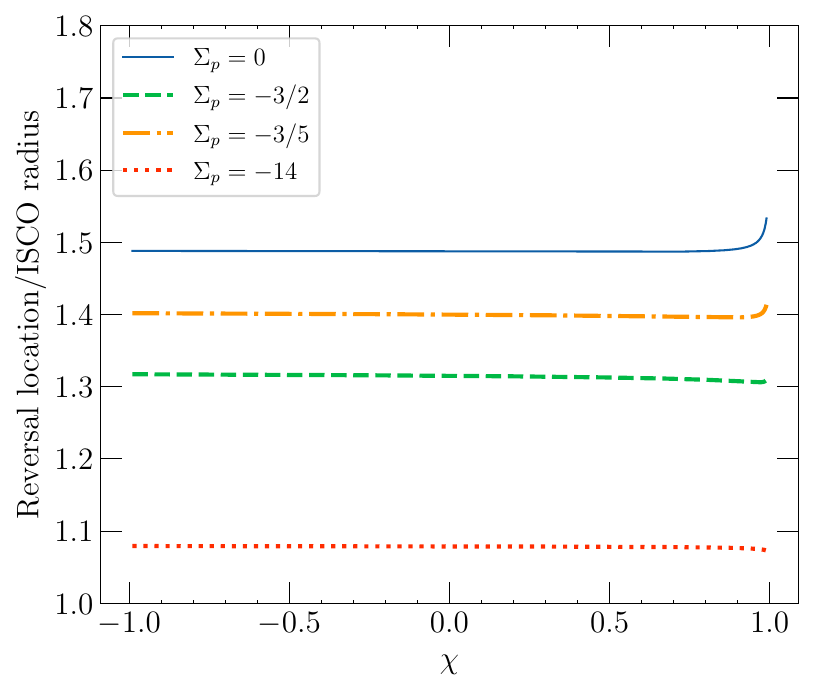}
    \caption{The dependence of the torque reversal location on the surface density gradient. Observe that the location approaches the ISCO for large and negative values of $\Sigma_{p}$ and the location roughly remains constant with respect to the ISCO as a function of the spin of the SMBH.}
    \label{fig:location-of-reversal-dependence-on-gradients}
\end{figure}
Our assumption that the inner and the outer cutoff parameters are the same allows us to simplify Eqs.~\eqref{eq:Ldot-Relativistic-expr-LR} and \eqref{eq:lindblad-resonance-total} to
\begin{align}\label{eq:lindblad-resonance-total-sim}
    \left<\dot{L}'_z\right>_{\mathrm{Lr,tot}} 
    &
    = 
    \frac{j_{\mathrm{max}}^2}{\Omega(p')}\left< \dot{E}' \right>_{\mathrm{Lr,+}}^{(1)} 
    \,.
\end{align}
From Eq.~\eqref{eq:Edot-Relativistic-expr-LR}, we see that $\left< \dot{E}' \right>_{\mathrm{Lr,+}}^{(1)}$ is the sum of two different terms.
The first term is proportional to $d\Sigma[p']/dp'$ and the second term is proportional to $\Sigma[p']$.
The coefficient of $d\Sigma[p']/dp'$ can be shown to be negative definite if one is outside the ISCO. Unless $d\Sigma[p']/dp'$ is large, this term is subdominant compared to the term proportional to $\Sigma(p')$.
The sign of the term proportional to $\Sigma(p')$ is controlled by 
\begin{align}\label{eq:G1-location}
    \mathcal{G}_{1}(p') \equiv -b_1\left[K_0(X) + b_0 K_1(X) \right]\left[K_0(X) + b_2 K_1(X) \right]
    \,.
\end{align}
In the Newtonian limit, $\mathcal{G}_{1}(p')<0$ leading to a net negative torque on the SCO.
In a Schwarzschild background spacetime, $\mathcal{G}_{1}(p')<0$ only if $p'>8.92$, signaling that there is a torque reversal around $p'\approx 8.92$ if the contribution from the term proportional to $d\Sigma[p']/dp'$ is not too large.

The location of the torque reversal, determined by ${\cal{G}}_1(p')=0$, was displayed as an orange dash-dotted curve in Fig.~\ref{fig:location-of-reversal} as a function of the SMBH spin.
In Fig.~\ref{fig:location-of-reversal-dependence-on-gradients}, we show how the reversal location depends on the gradient of $\Sigma$ by varying the parameter $\Sigma_{p}$. 
The orange dash-dotted curve $(\Sigma_p = -3/2)$ and the green dashed curves $(\Sigma_p=-3/5)$ show the torque reversal location for $\alpha$ and $\beta$ disks respectively.
Observe that the reversal location gradually approaches the ISCO as we decrease $\Sigma_{p}$, and once we reach $\Sigma_{p} =-14$ (red dashed curve), the reversal location is very close to the ISCO.
Hence, unless $\Sigma_{p}$ is large and negative, the torque reversal always occurs outside the ISCO. 
From the plot, we also see that the location is approximately insensitive to the spin of the SMBH, unless $\chi$ is very close to 1, as already hinted at when we discussed Fig.~\ref{fig:location-of-reversal}. 

\subsection{Comparison to Newtonian theory}\label{sec:comparison-to-Newt}
\begin{figure}
    \centering
    \includegraphics[width=1\linewidth]{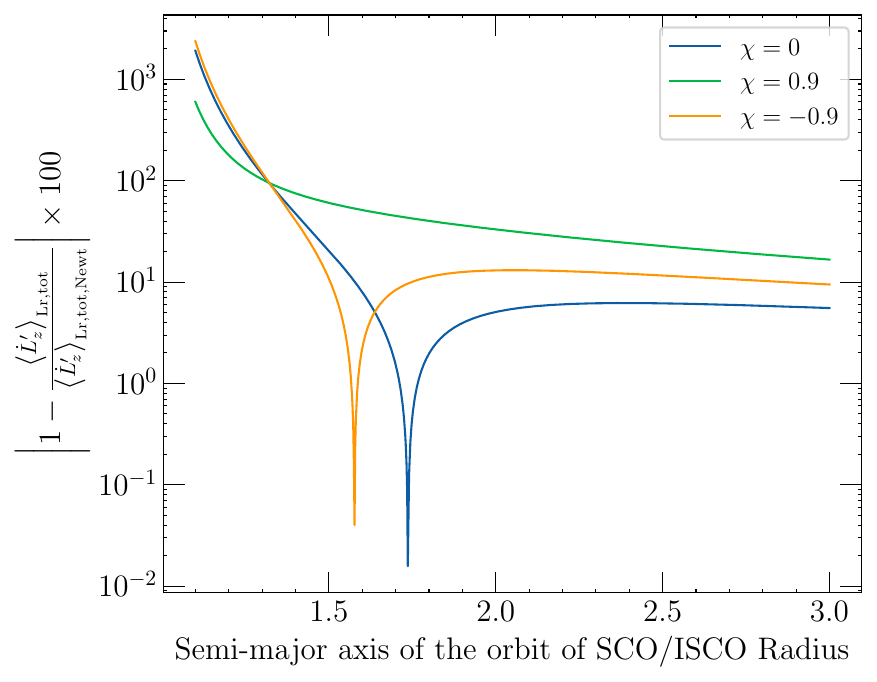}
    \caption{The percentage difference between the Newtonian and the relativistic torque formula for different values of the spin of the SMBH as a function of the location of the SCO from the ISCO.}
    \label{fig:diff-with-Newtonian}
\end{figure}
\begin{figure*}[thp]
    \centering
    \includegraphics[width=0.48\linewidth]{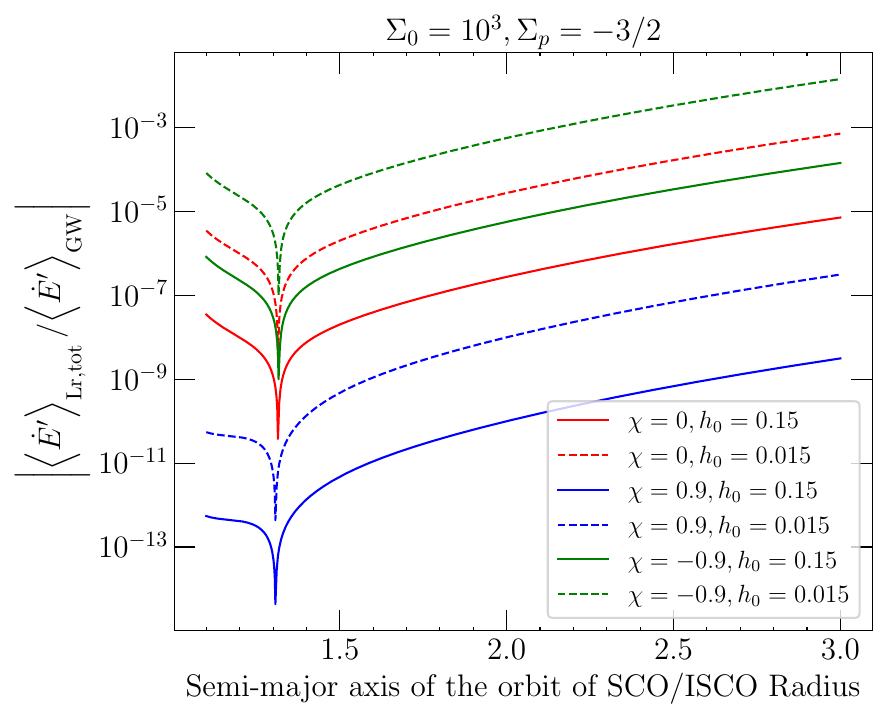}
    \includegraphics[width=0.48\linewidth]{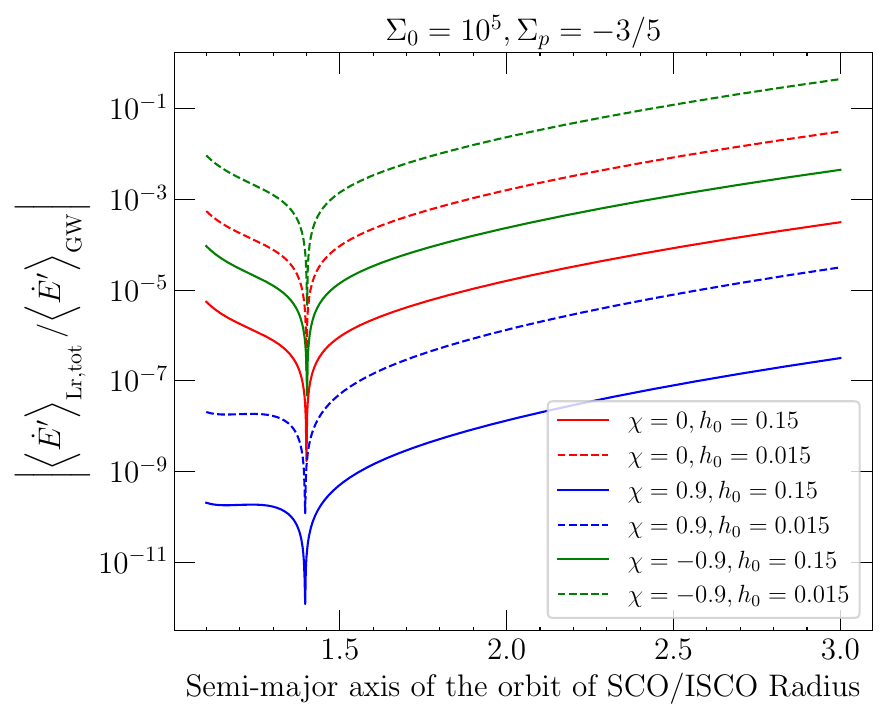}
    \caption{Ratio of the rate of energy exchange due to disk-CO interactions to gravitational wave loss to infinity for a few different values of the aspect ratio and spin of the SMBH. The left panel shows the ratio for $\Sigma_0 = 10^3$ and $\Sigma_p=-3/2$, which are $\alpha$ disk-like values.
    The right panel uses $\Sigma_0 = 10^5$ and $\Sigma_p=-3/5$, which are consistent with $\beta$ disk-like values.
    Observe that for both panels, the ratio decreases with the spin of the SMBH because the ISCO is closer to the horizon of the SMBH, and gravitational wave energy loss increases drastically as we approach the horizon.}
    \label{fig:disk-SCO-to-GW}
\end{figure*}
We now compare the relativistically accurate expression obtained in Eq.~\eqref{eq:lindblad-resonance-total-sim} to the Newtonian torque used in the literature~\cite{Ward-1986},
\begin{align}
    &\left<\dot{L}'_z\right>_{\mathrm{Lr,tot,Newt}}
    =
    \frac{j_{\mathrm{max}}^2}{\Omega(p')}\left< \dot{E}' \right>_{\mathrm{Lr,+,\mathrm{Newt}}}^{(1)}
    =
    \nonumber\\
    &
    -\frac{4M q^2 j_{\mathrm{max}}^2 \Sigma}{9 x^2\Omega(p')}
    \bigg\{
    2 \frac{d \log(\Sigma)}{d \log (p')}
    \bigg(2 K_0\left(\frac{2}{3} \right)+K_1\left(\frac{2}{3}\right)\bigg)^2
    \nonumber\\
    &+
    \bigg[7 K_0\left(\frac{2}{3} \right)+8 K_1\left(\frac{2}{3}\right)\bigg] \bigg[2 K_0\left(\frac{2}{3} \right)+K_1\left(\frac{2}{3}\right)\bigg]
    \bigg\}
    \,.
\end{align}
In Fig.~\ref{fig:diff-with-Newtonian}, we plot the percentage relative fractional difference between the Newtonian and the relativistic expressions as a function of the distance of the SCO orbit from the ISCO for a few different values of the SMBH spin with $\Sigma_{p}=-3/2$ (i.e.~an $\alpha$-disk like value).
Notice that the difference can be as high as 10-1000 \%, which indicates that one should definitely include the relativistic corrections to reliably model the torque induced on a SCO in an EMRI in regions close to the SMBH, which is precisely the regime of interest to gravitational-wave observations. In general, we see that spin tends to make the relativistic corrections to the torque stronger.

\subsection{Comparison to gravitational wave emission}\label{sec:comparison-to-GW}
The ratio of the rate of energy exchanged through the disk-SCO interaction to gravitational-wave emission has the characteristic scaling
\begin{align}
    \left|\frac{\left<\dot{E}'\right>_{\mathrm{Lr,tot}}}{\left<\dot{E}'\right>_{\mathrm{GW}}} \right|
    \propto
    \frac{M \Sigma_{0}}{h_0^2}
    \,.
\end{align}
In Fig.~\ref{fig:disk-SCO-to-GW}, we plot this ratio for a SMBH with mass $M = 10^6 M_{\odot}$ surrounded by an accretion disk with $\Sigma_{h} = 1$ and $\Sigma_{0} = 10^3$, $\Sigma_{p} = -3/2$ ($\alpha$ disk-like values) on the left panel and $\Sigma_{0} = 10^5$, $\Sigma_{p} = -3/5$ ($\beta$ disk-like values) on the right panel.
In both panels, we choose a few different values of SMBH spin and accretion disk aspect ratio.
For the gravitational wave flux, we use the analytical expressions from~\cite{Fujita_2015}, as coded in \texttt{Black Hole Perturbation Toolkit}~\cite{BHPToolkit}.
As we see from the figure, the value of the energy flux due to the disk-SCO interaction is always smaller than the loss of energy to gravitational-wave emission.
Therefore, the torque reversal \textit{cannot} cause a floating orbit, unless $\Sigma_{0}$ is extremely large.
We also see that, for the $\chi = 0.9$ case, the torque is smaller than the $\chi=0$ and $\chi=-0.9$ values; this is because the ISCO is very close to the horizon, and therefore, the energy loss due to gravitational wave emission is always significantly larger.
For a retrograde SCO orbit and a retrograde disk ($\chi=-0.9$), we see that the torque is only 1-3 orders of magnitude smaller than the gravitational-wave contribution, because the ISCO approaches $9M$ as $\chi \to -1$.
This suggests that for retrograde orbits, the disk-SCO interaction will have a significantly larger impact on the waveform.
Finally, although the ratio is very small, the large amount of cycles spent by an EMRI in the LISA band could help us disentangle the disk-SCO effect from the vacuum EMRI signal.
We refer the reader to Paper I for a heuristic dephasing estimate in a Schwarzschild background spacetime. For Bayesian and Fisher estimates using the Newtonian torque formula, see~\cite{Yunes:2011ws,Kocsis_2011,Speri_2023}.

\section{Conclusions}\label{sec:conclusions}
In this paper, we extended the results of Paper I by analyzing the relativistic treatment of disk-SCO interactions to a Kerr background spacetime.
For nearly circular and equatorial orbits in a Kerr spacetime, we provided a detailed analytic exploration of the effects of the spin on the torque due to Lindblad resonances.
Our two key results are the following. First, we provided analytical expressions for the power and torque induced on the SCO from the disk-SCO interaction in an EMRI [Eqs.~\eqref{eq:Lindblad-E-expr-actual} and \eqref{eq:Ldot-Relativistic-expr-LR}]. Second, we investigated the impact of the SMBH spin and the accretion disk surface density gradient on the torque reversal location [ Figs.~\ref{fig:location-of-reversal} and \ref{fig:location-of-reversal-dependence-on-gradients}]. We have found that the spin of the Kerr background tends to enhance the relativistic disk-SCO interaction, and this must be taken into account in future EMRI modeling of environmental effects.
Some of these results have been confirmed by an independent work~\cite{future-work-in-prep-Duque}, albeit numerically.

Given that we now have a qualitative understanding of how strong relativistic gravity impacts disk-SCO interactions in the close vicinity of a SMBH for nearly circular and equatorial orbits, future work must concentrate on modeling disk-SCO interactions in more generic eccentric and inclined orbits.
The techniques introduced in Paper I will readily generalize to such orbits. An interesting question for generic orbits in a Kerr spacetime is whether transient resonances on inclined orbits can interact and saturate due to accretion disk interactions.

A more significant issue is how sensitive the analysis of relativistic disk-SCO interactions is to the highly simplified disk model that we have used. In particular the disk is likely
to contain density fluctuations that can exert torques on the SCO, as well as exciting
its eccentricity and inclination~\cite{sun2025probingformationchannelsextreme}.
The response and structure of the disk will be mediated by
magnetized turbulence in the presence of radiative cooling.   And finally, accretion onto the
SCO will produce feedback in the disk that may also modify the torques.   
To explore these future directions one can broadly use two different methods to perform this task, drawing motivation from Newtonian theory.
First, one could use linear perturbation theory to model the interaction between the SCO and the disk~\cite{1978ApJ...222..850G,1979ApJ...233..857G,1993ApJ...419..155A,1993Icar..102..150K,fairbairn2025pushinglimitseccentricityplanetdisc}.
This approach should be sufficient to understand the impact of Lindblad resonances. 
However, understanding the corotation resonances may require a more direct numerical approach to the problem.
Another interesting avenue in fluid modeling is how magnetic fields and stochastic torques could impact the disk-SCO interactions. Understanding this problem requires modeling the relativistic magnetohydrodynamical nature of the fluid disk, which could have a rich dynamical structure.

Finally, it would be straightforward to extend the approach introduced in Paper I to model intermediate mass ratio inspirals. However, depending on the mass ratio, the SCO in such inspirals could open a gap in the disk, leading to complicated gap-opening dynamics that can only be modeled numerically~\cite{Paardekooper:2023}.

\acknowledgements
We thank Yuri Levin for bringing our attention to the possibility that the accretion onto the SCO could potentially cause a torque reversal.
A.~H.~K.~R.~and N.~Y.~acknowledge support from the Simons Foundation through Award No. 896696, the NSF through Grant No. PHY-2207650, NASA through Grant No. 80NSSC22K0806 and the Simons Foundation International through Award No. SFI-MPS-BH-00012593-01.  C.~F.~G. acknowledges support from the NSF through AST 20-34306.  Parts of this work were completed at the Aspen Center for Physics, which is supported by National Science Foundation grant PHY-2210452.
\appendix
\section{Geodesics in the equatorial plane}\label{appendix:orbital-mechanics-Kerr}
In this appendix, we review the motion of geodesics in the Kerr spacetime confined to the equatorial plane. 
The material presented here is standard material reviewed in many references, see e.g.~\cite{Chandrasekar-BH-Book,Pound_2021}.
We include this discussion for completeness and to setup our notation. 

The line element of the Kerr spacetime in Boyer-Lindquist coordinate system $(t,r,\theta,\phi)$ is given by
\begin{align}\label{eq:line-element-Kerr}
    ds^2 &= - \left[1- \frac{2M r}{\Sigma} \right] dt^2
    -
    \frac{4 a M r \sin^2(\theta)}{\Sigma} dt d\phi
    +
    \frac{\Sigma}{\Delta} dr^2
    \nonumber\\
    &+
    \Sigma d \theta^2
    +
    \left[
    \Delta + \frac{2Mr (r^2 + a^2)}{\Sigma} 
    \right]
    \sin^2(\theta) d \phi^2
    \,,
\end{align}
where $M$ is the mass of the black hole, $\chi := a/M$ is the dimensionless spin, $\Sigma = r^2 + a^2 \cos(\theta)^2$ and $\Delta = r^2 - 2 M r + a^2$.
Geodesic motion of a particle of mass $m$ moving on a worldline $z^{\mu}(t)$ in the Kerr spacetime is generated by the specific Hamiltonian
\begin{align}\label{eq:hamiltonian-geodesic-motion}
    \mathscr{H} = -p_{t}
\end{align}
where $p^{\mu} = d z^{\mu}/d\tau$ is the specific four momentum and $p_{\mu} = g_{\mu \nu} p^{\mu}$.
We can obtain an explicit expression for the specific Hamiltonian from the normalization condition
\begin{align}
    p_{t} p_{t} g^{tt} + 2 g^{ti}  p_{t} p_{i} + p_{i} p_{j} g^{ij} = -1\,,
\end{align}
which yields
\begin{align}
    p_{t} &= \frac{-g^{ti} p_i + \sqrt{(g^{ti}p_i)^2 - g^{tt} (g^{ij} p_{i}p_{j} + 1)}}{g^{tt}} \,.
\end{align}
One can simplify this equation using the line element [Eq.~\eqref{eq:line-element-Kerr}] to obtain an explicit expression for the Hamiltonian in terms of coordinates [Eq.~\eqref{eq:hamiltonian-geodesic-motion}].

We now restrict our attention to equatorial orbits $(\theta = \pi/2, p_{\theta} = 0)$ and obtain the action angle variables of the system using the Hamilton-Jacobi equation
\begin{align}\label{eq:Hamilton-Jacobi-Equation}
    \mathscr{H}\left[x^{\nu} , \frac{\partial \mathscr{S}}{\partial x^{\nu}}\right]
    +
    \frac{\partial \mathscr{S}}{\partial t}
    &=0\,,
\end{align}
where $\mathscr{S}$ is Hamilton's principal function.
To solve Eq.~\eqref{eq:Hamilton-Jacobi-Equation}, we use the ansatz
\begin{align}
    \mathscr{S} = - \mathscr{E} t + \mathscr{W}(x)  \,,
\end{align}
where $\mathscr{W}(x)$ is called Hamilton's characteristic function and $\mathscr{E}$ is the specific energy of the system. 
Standard techniques in Hamiltonian mechanics~\cite{Goldstein-Safko-book,Chandrasekar-BH-Book,Hinderer_2008} guides us to use the following ansatz for $\mathscr{W}$:
\begin{align}\label{eq:Hamiltons-characteristic-ansatz}
    \mathscr{W} &= \mathscr{L}_{z} \; \phi + W_{R}(r)  \,,
\end{align}
where $\mathscr{L}_z$ is the angular-momentum in the z-direction.
The function $W_R$ is determined by solving the Hamilton-Jacobi equation
\begin{align}
    \left(\frac{d W_R}{dr}\right)^2
    &=
    \frac{V_r(r)^2}{\Delta^2}\,,
\end{align}
where $V_r(r)$ is defined as~\cite{Hinderer_2008}
\begin{align}\label{eq:WR-sol}
    V_r &\equiv 
    \left[\mathscr{E}\left(a^2+r^2\right)
    -a \mathscr{L}_z\right]^2
    -\Delta (r) \left[(\mathscr{L}_z-a \mathscr{E})^2+r^2\right]
    \,.
\end{align}
Given a solution to the Hamilton-Jacobi equation, we can obtain the canonical momentum of the system using
\begin{align}
    p_{a} = \frac{\partial \mathscr{S}}{\partial z^{a}}\,.
\end{align}
Substituting this in Eqs.~\eqref{eq:Hamiltons-characteristic-ansatz} and \eqref{eq:WR-sol}, we see that
\begin{subequations}
\begin{align}
    p_{r} &= \frac{\partial W_R}{\partial r} = \frac{\sqrt{V_r}}{\Delta}\,,\qquad
    p_{\phi} = \mathscr{L}_{z} \,.
\end{align}
\end{subequations}
We are now in a position to compute the action variables $(J_{r}, J_{\phi})$ of the system 
\begin{subequations}
\begin{align}
    \label{eq:Jr-v1}
    J_{r}
    &= \frac{1}{2 \pi} \oint \frac{dW_R}{dr} dr
    =
    \frac{1}{\pi} \int_{r_{\mathrm{min}}}^{r_{\mathrm{max}}} \frac{\sqrt{V_r}}{\Delta} dr
    \,,\\
    \label{eq:Jphi-v1}
    J_{\phi} &= \mathscr{L}_z \,.
\end{align}
\end{subequations}
where $r_{\mathrm{min}}$ and $r_{\mathrm{max}}$ are the two largest roots of $V_r$.
The angle variables $q_{\alpha}$ and the frequencies $\nu_{\alpha}$ of the system are obtained from $\mathscr{S}$ and $\mathscr{H}\left(J_a\right)$ using
\begin{subequations}
\begin{align}
    \label{eq:angle-q-vars}
    q_{a} &= \frac{\partial \mathscr{W}}{\partial J_{a}}\,, \\
    \label{eq:nu-definition}
    \nu_{a} &= \frac{\partial \mathscr{H}(J)}{\partial J_{a}} = \frac{\partial \mathscr{E}\left[J_{r}, J_{\phi}\right]}{\partial J_{a}}
    \,.
\end{align}
\end{subequations}
One can obtain general expressions for the frequencies and the orbital elements of the system using the techniques described in~\cite{Schmidt_2002,Fujita_2009}.
For the analysis presented in this paper, it is sufficient to use a small-eccentricity expansion.

For bound orbits, we adopt the quasi-Keplerian parameterization
\begin{align}
    r &= \frac{pM}{1 + e \cos(\chi_{\mathrm{orb}})}\,.
\end{align}
where $e$ is the orbital eccentricity and $\chi_{\mathrm{orb}}$ is the relativistic anomaly, which should not be confused with the dimensionless spin $\chi$. Using this parametrization, we can factor $V_r$ as
\begin{align}
    V_r &= \left(1- \mathscr{E}^2\right)(r_a - r)(r - r_p)(r-r_3) r
\end{align}
where
\begin{align}
    r_a &= \frac{pM}{1-e}\,, \quad
    r_p = \frac{pM}{1+e} \,, \quad
    r_3 = \frac{2M}{1-\mathscr{E}^2} - (r_a + r_p) \,.
\end{align}
The quasi-Keplerian parameterization can be used to obtain expressions for the orbital elements and the action-angle variables.
A detailed account of this procedure is provided in~\cite{Schmidt_2002,Fujita_2009}.
Here, we are only interested in obtaining expressions in a small eccentricity expansion. 
We now schematically describe how this is achieved, and we present the analytical expressions for the orbital elements in Appendix~\ref{appendix:small-eccentricity}.

Using the fact that $V_r = 0$ at $r_a$ and $r_p$, we can obtain expressions for $\mathscr{E}$ and $\mathscr{L}_z$ (see Appendix B of~\cite{Schmidt_2002}).
Finally, we use the quasi-Keplerian parameterization to integrate Eq.~\eqref{eq:Jr-v1} and obtain expression for $J_r$ in a small eccentricity expansion.
To obtain expressions for $q_r$ and $q_{\phi}$, we need the evolution of the coordinates of the worldline.
This can be obtained using the quasi-Keplerian parameterization with the definition $p_{\mu} = g_{\mu \nu} d z^{\nu}/d\tau$ 
\begin{subequations}\label{eq:orbital-evolution-equations}
\begin{align}
    \frac{dt}{d\chi_{\mathrm{orb}}} &= \frac{r \left(-2 a \mathscr{L}_z M+\mathscr{E} r^3+a^2 \mathscr{E} (2 M+r)\right) r'}{p_r \Delta ^2}\,,\\
    \frac{d\tau}{d\chi_{\mathrm{orb}}} &= \frac{r^2 r'}{p_r \Delta}\,,\\
    \frac{d\phi}{d\chi_{\mathrm{orb}}} &= \frac{r (2 a \mathscr{E} M+\mathscr{L}_z (-2 M+r)) r'}{p_r \Delta ^2}\,,
\end{align}
\end{subequations}
where $r' = dr/d\chi_{\mathrm{orb}}$.
From Eqs.~\eqref{eq:Hamiltons-characteristic-ansatz} and ~\eqref{eq:angle-q-vars}, we see that
\begin{subequations}\label{eq:angle-q-exprs-1}
\begin{align}
    q_r &= \frac{\partial W_R}{\partial \mathscr{E}} \frac{\partial \mathscr{E}}{\partial J_r}\,,\\
    q_{\phi} &= \phi + 
    \frac{\partial W_R}{\partial \mathscr{E}} \frac{\partial \mathscr{E}}{\partial \mathscr{L}_z}
    +
    \frac{\partial W_R}{\partial \mathscr{L}_z} 
    \,.
\end{align}
\end{subequations}
We can obtain explicit expressions for the quantities appearing above by differentiating under the integral sign.
Finally, we can use Eq.~\eqref{eq:orbital-evolution-equations} to obtain explicit expressions for $q_r$ and $q_{\phi}$.
One can follow the same approach and obtain expressions for the frequencies of the system from the above equations.

Using a contact transformation, we define a new set of action angle coordinates $\left(\left\{\Lambda,\lambda\right\},\left\{P,\mathscr{p} \right\}\right)$ via
\begin{subequations}\label{eq:modified-Delaunay-variables}
\begin{align}
    \Lambda &\equiv J_r + J_{\phi} \,, &\lambda&\equiv q_{\phi} \,, &\nu_{\lambda}&\equiv \nu_{\phi} \,,\\
    P &\equiv J_r \,, &\mathscr{p}&= -\varpi \equiv q_r - q_{\phi} \,, &\nu_{\mathscr{p}}&\equiv \nu_r - \nu_{\phi} \,.
\end{align}
\end{subequations}
One can convert $\chi$ and $\omega \equiv \phi(\chi=0)$ to $\lambda$ and $\varpi$ by inverting Eq.~\eqref{eq:angle-q-exprs-1} and using Eq.~\eqref{eq:modified-Delaunay-variables}.
\section{Small eccentricity expansions}\label{appendix:small-eccentricity}
The small eccentricity expansion of $r$ and $\phi$ are given by
\begin{subequations}\label{eq:orb-elements-expansion-small-eccentricity}
\begin{align}
    &r = M p-e M p \cos (\lambda -\varpi )
    + \mathcal{O}(e^2) 
    \,,\\
    &\phi = \lambda 
    \nonumber\\
    &+ 
    \frac{2 e x_0^{12} \left(2 \chi+x_0^6-3 x_0^2\right) \sin (\lambda -\varpi )}{\left(\chi+x_0^6\right) \sqrt{8 \chi x_0^2-3 \chi^2+x_0^8-6 x_0^4} \left(\chi^2+x_0^8-2 x_0^4\right)}
    \nonumber\\
    &+\mathcal{O}(e^2)\,,
\end{align}
\end{subequations}
where $x_0 = p^{1/4}$. Expanding the action variables, one finds
\begin{subequations}
\begin{align}
    \Lambda
    &=
    \frac{M \left(-2 \chi x_0^2+\chi^2+x_0^8\right)}{x_0^3 \sqrt{2 \chi+x_0^6-3 x_0^2}}
    +
    \mathcal{O}(e^2) \,,\\
    P
    &=
    \frac{e^2 M x_0^9 \sqrt{-3 \chi ^2+8 \chi  x_0^2+x_0^8-6 x_0^4}}{2 \sqrt{2 \chi +x_0^6-3 x_0^2} \left(\chi ^2+x_0^8-2 x_0^4\right)}
    +
    \mathcal{O}(e^3)
    \,.
\end{align}
\end{subequations}
Expansion of the frequencies of the system are given by
\begin{subequations}
\begin{align}
    \nu_{\lambda}
    &=
    \frac{1}{M \chi +M x_0^6}
    +
    \mathcal{O}(e^2)
    \,,\\
    \nu_{\varpi}
    &=
    \frac{x_0^4-\sqrt{-3 \chi ^2+8 \chi  x_0^2+x_0^8-6 x_0^4}}{M \chi  x_0^4+M x_0^{10}}
    +
    \mathcal{O}(e^2)
    \,.
\end{align}
\end{subequations}
\section{Coefficients $b_{1}$ and $b_{2}$}\label{appendix:b-12}
The coefficients $b_1$ and $b_2$ are given by
\begin{subequations}
\begin{align}
    b_{1} &= \frac{b_{1,n}}{b_{1,d}} \,,\\
    b_{2} &= \frac{b_{2,n}}{b_{2,d}}
    \left(\frac{x \sqrt{2 \chi +x^6-3 x^2}}{\sqrt{-3 \chi ^2+x^8-6 x^4+8 \chi  x^2}}\right)^{-1}
    \,.
\end{align}
\end{subequations}
where
\begin{widetext}
\begin{subequations}
\begin{align}
    &b_{1,n} = 
    -x^{16} 
    \bigg\{
    72 \chi ^7-21 x^{30}+331 x^{26}-360 \chi  x^{24}+\left(113 \chi ^2-1626\right) x^{22}+2880 \chi  x^{20}-27 \left(55 \chi ^2-124\right) x^{18}
    \nonumber\\
    &-24 \chi  \left(3 \chi ^2+340\right) x^{16}+3 \left(59 \chi ^4+2312 \chi ^2-792\right) x^{14}-192 \chi  \left(11 \chi ^2-36\right) x^{12}+\left(309 \chi ^4-6900 \chi ^2\right) x^{10}
    \nonumber\\
    &-8 \chi ^3 \left(51 \chi ^2-220\right) x^8+9 \chi ^4 \left(19 \chi ^2+130\right) x^6-512 \chi ^5 x^4-147 \chi ^6 x^2
    \bigg\}
    \left(2 \chi +x^6-3 x^2\right)^3
    \,,\\
    &b_{1,d} = 
    \left(\chi +x^6\right)^3 \left(-3 \chi ^2+x^8-6 x^4+8 \chi  x^2\right)^2 \left(\chi ^2+x^8-2 x^4\right)^4
    \,,\\
    &b_{2,n} = 
    2 \bigg[63 \chi ^7+\chi ^5 \left(63 x^4-236\right) x^4-\chi ^3 \left(15 x^8+96 x^4+244\right) x^8+45 \chi ^6 \left(x^4-4\right) x^2
    \nonumber\\
    &+3 \chi  \left(-69 x^{12}+548 x^8-1380 x^4+1056\right) x^{12}+15 \chi ^4 \left(6 x^8-38 x^4+75\right) x^6+\chi ^2 \left(65 x^{12}-864 x^8+2946 x^4-2520\right) x^{10}
    \nonumber\\
    &+\left(-12 x^{16}+190 x^{12}-927 x^8+1800 x^4-1188\right) x^{14}\bigg]
    \,,\\
    &b_{2,d} = 
    72 \chi ^7-8 \chi ^5 \left(51 x^4+64\right) x^4+24 \chi  \left(288-5 x^4 \left(3 \left(x^4-8\right) x^4+68\right)\right) x^{12}-8 \chi ^3 \left(9 x^8+264 x^4-220\right) x^8
    \nonumber\\
    &+3 \chi ^6 \left(57 x^4-49\right) x^2+3 \chi ^4 \left(59 x^8+103 x^4+390\right) x^6+\chi ^2 \left(113 x^{12}-1485 x^8+6936 x^4-6900\right) x^{10}
    \nonumber\\
    &+\left(-21 x^{16}+331 x^{12}-1626 x^8+3348 x^4-2376\right) x^{14}
    \,.
\end{align}
\end{subequations}
\end{widetext}
These coefficients, as well as the ones that appear in Eq.~\eqref{eq:dot-E-L-Lr-eqns}, are provided in the supplementary \texttt{MATHEMATICA} file.
\bibliography{ref}
\end{document}